\documentclass[aps,prl,twocolumn,superscriptaddress]{revtex4-1}

\usepackage{graphics}
\usepackage{epsfig}

\begin{document}

\preprint{(to be submitted to Phys. Rev. Lett.)}

\title{{\it In situ} X-ray photoelectron spectroscopy of model catalysts: At the edge of the gap}

\author{S. Blomberg}
\email[]{sara.blomberg@sljus.lu.se}
\affiliation{Div.~of Synchrotron Radiation Research, Lund University, Box 118, SE-221 00 Lund, Sweden}
\author{M.J. Hoffmann}
\affiliation{Dept.~Chemie, Technische Universit{\"a}t M{\"u}nchen, Lichtenbergstr. 4, D-85747 Garching, Germany}
\author{J. Gustafson}
\affiliation{Div.~of Synchrotron Radiation Research, Lund University, Box 118, SE-221 00 Lund, Sweden}
\author{N. M. Martin}
\affiliation{Div.~of Synchrotron Radiation Research, Lund University, Box 118, SE-221 00 Lund, Sweden}
\author{V. R. Fernandes}
\affiliation{Dept.~of Physics, Norwegian Univ. of Sci. and Techn., NO-7491 Trondheim, Norway}
\author{A. Borg}
\affiliation{Dept.~of Physics, Norwegian Univ. of Sci. and Techn., NO-7491 Trondheim, Norway}
\author{Z. Liu}
\affiliation{ALS, Lawrence Berkeley National Laboratory, Berkeley, CA 94720, USA}
\author{R. Chang}
\affiliation{ALS, Lawrence Berkeley National Laboratory, Berkeley, CA 94720, USA}
\author{S. Matera}
\affiliation{Dept.~Chemie, Technische Universit{\"a}t M{\"u}nchen, Lichtenbergstr. 4, D-85747 Garching, Germany}
\author{K. Reuter}
\affiliation{Dept.~Chemie, Technische Universit{\"a}t M{\"u}nchen, Lichtenbergstr. 4, D-85747 Garching, Germany}
\author{E. Lundgren}
\affiliation{Div.~of Synchrotron Radiation Research, Lund University, Box 118, SE-221 00 Lund, Sweden}

\date{\today}

\begin{abstract}
We present a High-Pressure X-ray Photoelectron Spectroscopy (HP-XPS) and first-principles kinetic Monte Carlo study addressing the nature of the active surface in CO oxidation over Pd(100). Simultaneously measuring the chemical composition at the surface and in the near-surface gas-phase, we reveal both O-covered pristine Pd(100) and a surface oxide as stable, highly active phases in the near-ambient regime accessible to HP-XPS. Surprisingly, no adsorbed CO can be detected during high CO$_2$ production rates, which can be
explained by a combination of a remarkably short residence time of the CO molecule on the surface and mass-transfer limitations in the present set-up.
\end{abstract}
\maketitle

Understanding the detailed structure and nature of the active site is a central paradigm in modern molecular-level catalysis. For transition metal (TM) based heterogeneous catalysts this has motivated extensive studies of low-index single-crystal model catalysts, initially under controlled ultra-high vacuum (UHV) conditions \cite{ertl} and increasingly at higher pressures \cite{Lundgren}. Not withstanding, despite significant advances in {\em in situ} methods for surface characterization \cite{MRSStierle}, even qualitative structural and compositional properties remain to date surprisingly unclear for the technological near-ambient regime.

There is little doubt that late TMs like Pd oxidize under ambient oxygen pressures, and for low-index surfaces the evolving O-phases are well understood. At Pd(100), these are for instance two ordered O overlayers in UHV with $p(2 \times 2)$ and $c(2 \times 2)$ periodicity at 0.25 and 0.5\,monolayer (ML, defined as number of O atoms per Pd surface atom) coverage, respectively \cite{Altman1,LundgrenS5}. If the pressure is increased above $1 \times 10^{-6}$~Torr and the temperature is kept at 300$^\circ$C, oxidation proceeds to a well-ordered ($\sqrt{5}\times\sqrt{5}$)R27$^{\circ}$ (henceforth $\sqrt{5}$ for brevity) surface oxide structure, corresponding to a single PdO(101) plane on top of Pd(100) \cite{LundgrenS5,Kostelnik}. At pressures beyond 1\,Torr and 300$^\circ$C bulk PdO is formed \cite{lundgren04,Rasmus}.

What remains unclear, though, is whether or to what degree the presence of the other reactant, CO in the case of CO oxidation, inhibits oxide formation under steady-state operation. So far, well controlled semi-realistic CO oxidation studies over Pd(100) have been performed {\it in situ} using Scanning Tunneling Microscopy (STM) \cite{HendriksenPdSTM}, Polarization-Modulation Infrared Absorption Spectroscopy (PM-IRAS) \cite{Goodman} and Surface X-Ray Diffraction (SXRD) \cite{Richard}, as well as Density-Functional Theory (DFT) calculations \cite{Rogal}. Despite the alleged simplicity of the reaction, the interpretation of this data with respect to the nature of the active phase under technological ambient gas-phase conditions is still controversial \cite{RichardII}. Part of the difficulties lies in the high reactivity of the unselective CO oxidation reaction, which gives rise to significant mass-transfer limitations (MTLs) and makes the results heavily sensitive to the different reactors used (see below). A significant part of the disagreement, however, also arises from differences between the experimental techniques. While PM-IRAS probes one of the reactants (CO) only, STM and SXRD are sensitive to the surface structure and morphology of the substrate. In contrast, X-ray Photoelectron Spectroscopy (XPS) enables detection of adsorbates (CO and oxygen) and the substrate simultaneously, and in the case of high-pressure XPS (HP-XPS), also the gas-phase in the immediate vicinity of the model catalyst. This comprehensiveness of the information provided is a feature unique to this technique, with great potential for breakthrough discoveries in the surface catalytic context.

In an effort to further close the pressure gap between atomic-scale studies in UHV and real catalysis at ambient conditions we illustrate this with a HP-XPS study of CO oxidation over Pd(100) covering the entire pressure range up to 1 Torr. Supported by detailed first-principles kinetic Monte Carlo (1p-kMC) modeling the obtained data show exclusively that both O-covered pristine Pd(100) and the $\sqrt{5}$ surface oxide are highly active phases in the near-ambient regime. Which phase gets stabilized depends sensitively on temperature, total pressure, feed stoichiometry, i.e. the CO:O$_2$ partial pressure ratio, and due to MTLs also the macroscopic flow profile in the employed reactor. Our interpretation favors the presence of the surface oxide at technological conditions, but validation of this hypothesis will require extension of existing {\em in situ} techniques to ambient pressures and new reactor setups to control the severe MTLs clearly identified in this work.

The HP-XPS measurements were performed at the Molecular Science beamline 9.3.2 at the ALS in Berkeley \cite{Zhi,chang}. XPS measurements {\it in situ} in gas pressures up to 1 Torr can be performed. The Pd 3$d_{5/2}$, C 1$s$ and O 1$s$ core levels were recorded with photon energies of 435, 435 and 650 eV, respectively. Interpretation of the measurements was aided by 1p-kMC simulations focusing exclusively on the metal Pd(100) surface. With further details provided in the supplementary material, the model considers adsorption, desorption, diffusion and reaction processes at a Pd(100) lattice, with O adatoms occupying the fourfold hollow and CO occupying bridge sites. Repulsive lateral interactions are described through nearest-neighbor site-blocking rules, and all kinetic parameters entering the simulations were determined by supercell geometry DFT calculations \cite{CASTEP}, using the semi-local PBE functional \cite{PBE}. Steady-state catalytic activity was evaluated for given partial pressures and increasing temperatures. Observing a sharp increase in activity over a narrow temperature range  we define the ''activation temperature'' (see below) as the temperature corresponding to the inflection point of the activity increase.

\begin{figure}
\includegraphics[width=7.8cm]{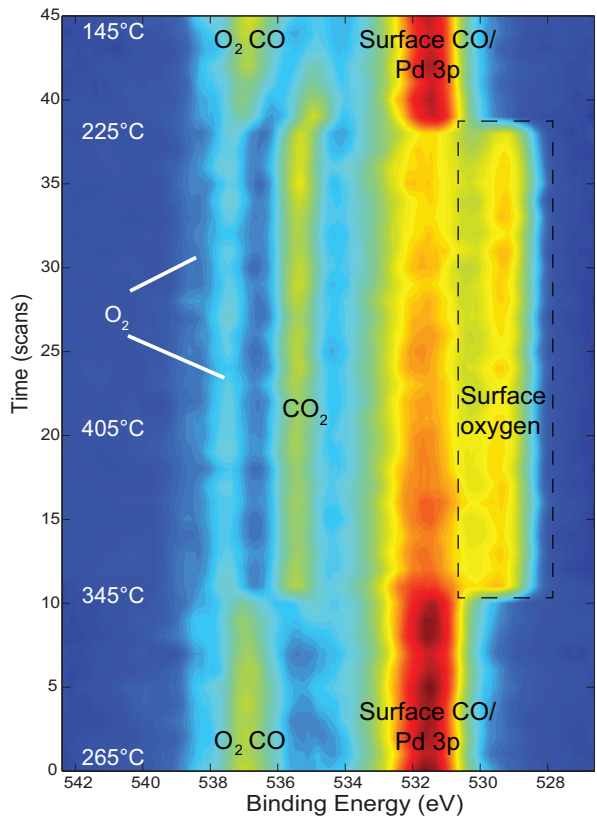}
\caption{\label{live} (Color online) O 1$s$ region during CO oxidation in a gas mixture of 0.25 Torr CO and 0.25 Torr O$_2$. The temperature  of the Pd(100) (shown to the left in the figure)  was ramped up and down during the measurement.}
\end{figure}

We start illustrating the insights provided by the {\em in situ} XPS measurements with the continuous O 1$s$ scans compiled in Fig.~\ref{live}. The chamber was filled with 0.25 Torr CO and 0.25 Torr O$_2$ and the temperature was ramped from 265$^\circ$C up to 405$^\circ$C and then back down to 145$^\circ$C. Under these conditions, the O 1$s$ region reveals the surface adsorbates, the phase of the substrate, as well as the composition of the gas above the surface, i.e. we may follow the surface structure and reactivity simultaneously. Starting at the bottom of the figure, the spectra show two major peaks corresponding to CO adsorbed on the surface, which unfortunately coincide with Pd 3$p$, as well as CO and O$_2$ in the gas-phase. The gas-phase peaks are difficult to resolve from this figure, but consist of two O$_2$ related components at 537.5 and 538.6 eV \cite{Siegbahn} and one CO component at 536.4 eV. The absence of a CO$_2$ peak shows that the reactivity at this point is low. As the temperature is increased to 345$^\circ$C, the CO$_2$ peak (535.5 eV) suddenly dominates the gas-phase region completely. There is still some oxygen, but the gas-phase peak of the CO minority species is gone. This shows that the sample has suddenly become so active that almost all the CO near the surface is converted into CO$_2$, i.e. the measurements clearly suffer from MTLs (see below). Simultaneously, the peak corresponding to adsorbed CO is replaced by one corresponding to adsorbed O, i.e. the surface coverage changes from CO-rich to O-rich. We do not, however, see a split of the surface oxygen peak that would have been a fingerprint of (surface) oxide formation \cite{LundgrenS5}. Further heating does not change the spectrum significantly. In the cooling process, the active phase is present until a temperature of 225$^\circ$C, when the CO reclaims the surface and the activity is turned off.

\begin{figure*}
\includegraphics[width=14cm]{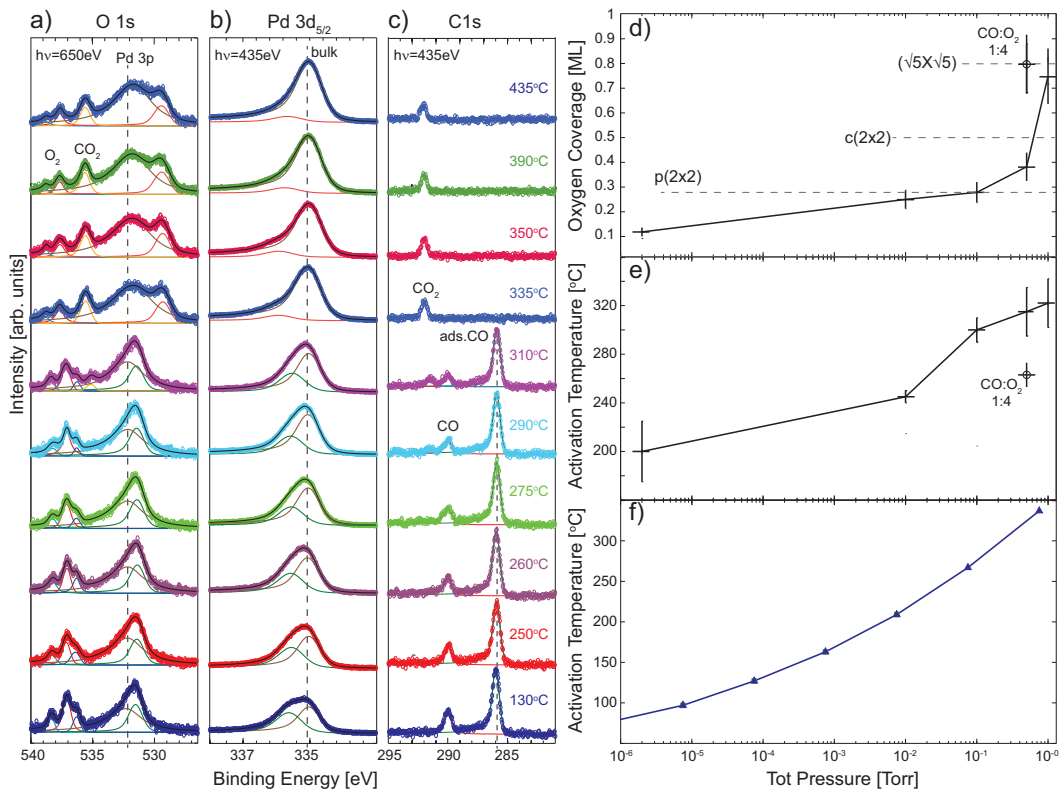}%
\caption{\label{250-250fit} (Color online) CO oxidation in a gas mixture of 0.25 Torr CO and 0.25 Torr O$_2$ showing the binding energy regions of a) O 1$s$, b) Pd 3$d$ and c) C 1$s$. d) Derived oxygen coverage and e) activation temperature for a CO:O$_2$ ratio of 1:1 and increasing total pressure. One measuring point from a CO:O$_2$ ratio of 1:4 is also included. f) Calculated 1p-kMC activation temperature for 1:1 CO:O$_2$ ratio and increasing total pressure.}
\end{figure*}

The resulting picture of a low-temperature CO-poisoned and a high-temperature active metallic state with chemisorbed O coverage is fully consistent with the conclusions derived from vibrational spectroscopy by Gao {\em et al.} for the same gas-phase ratios \cite{Goodman}. However, it was speculated that the latter active phase is of a transient nature due to the slow built-up of MTL-induced pressure gradients in the reactor \cite{RichardII}. To address this, we repeated the experiment with a finer step-wise temperature profile, and extending the measurements to the Pd 3$d$ and C 1$s$ region. The behavior of the O 1$s$ level (Fig.~\ref{250-250fit}a) is similar to Fig.~\ref{live} with the first signs of CO$_2$ production appearing at around 310$^\circ$C and the activation temperature at 335$^\circ$C. While the different heating speed has thus a slight effect on the observed activation temperature, a transient nature of the active phase can be excluded from the measured C 1$s$ level (Fig.~\ref{250-250fit}c). Below the activation temperature we observe gas-phase CO (289.9 eV) together with adsorbed CO in bridge sites (285.9 eV) \cite{Andersen}. At intermediate temperatures (310$^\circ$C), both CO and CO$_2$ can be detected in the gas-phase, and above the activation temperature only a single peak corresponding to CO$_2$ remains. The minor amount of CO still present in the small gas-phase volume to which we are sensitive is below the detection limit and can therefore not be observed in the spectra. Simultaneously, the Pd 3$d$ (Fig.~\ref{250-250fit}b) shows a slight shift towards lower binding energy as would be expected for adsorbed CO being replaced by O \cite{Andersen}. Above the activation temperature the mass transfer limited profile with very small amount of CO in the near-surface gas-phase is thus fully established in our measurements, with the active phase being metallic Pd(100) with chemisorbed O.

In a next step we continued the measurements at 1:1 CO:O$_2$ partial pressure ratio at different total pressures. The two-state behavior with defined activation temperature was always the same as the one just described, and we summarize in Fig.~\ref{250-250fit}e) the derived variation of the activation temperature with pressure. Even at the highest pressure attainable with the present {\em in situ} XPS setup, 1 Torr, we never observed a clear signature of surface oxide formation in the O 1$s$ spectrum. This is consistent with the $\pm 10$\% estimate of the O coverage above the activation temperature obtained from the ratios of the areas underneath the O 1\textit{s} and Pd 3\textit{p} levels against reference spectra of known oxygen structures prepared under UHV conditions \cite{LundgrenS5}. As apparent from Fig.~\ref{250-250fit}d), apart from the measurement at 1 Torr, this coverage remains at the level anticipated for the formation of $p(2 \times 2)$ or $c(2 \times 2)$ O overlayers. The conclusion that the active phase above activation temperature corresponds in the measured pressure range to O-covered metallic Pd(100) receives further support from our 1p-kMC simulations restricted to this metallic state. These simulations perfectly reproduce the two-state behavior with CO-poisoned low-temperature and O-covered active high-temperature state. The resulting activation temperature as a function of total pressure is shown in Fig.~\ref{250-250fit}f) and compares qualitatively well with the experimental data, considering the typical $\pm 100$$^\circ$C uncertainty due to the underlying semi-local DFT energetics \cite{reuter2004,SuppMat}.

\begin{figure*}[hbtp]
\begin{center}
\includegraphics[width=14cm]{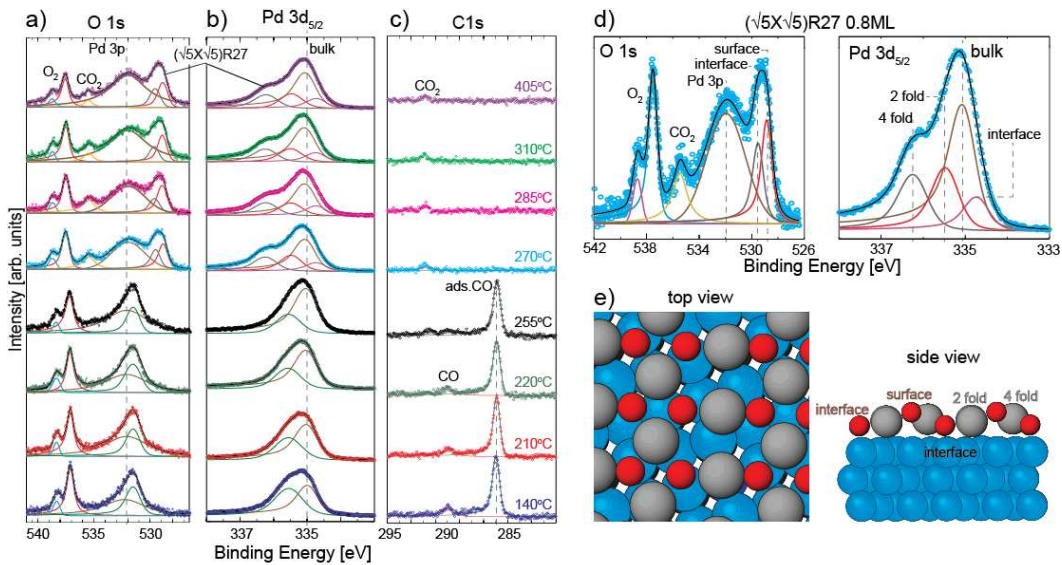}
\caption{\label{100-400fit} (Color online) CO oxidation at 0.5 Torr and a 1:4 CO:O$_2$ partial pressure ratio. Shown are the binding energy regions of a) O 1$s$, b) Pd 3$d$ and c) C 1$s$. d) A more detailed view of the O 1$s$ and Pd 3$d$ spectra directly above activation temperature. e) Structural model of the ($\sqrt{5}\times\sqrt{5}$) surface oxide on Pd(100) according to Ref. \onlinecite{LundgrenS5, Kostelnik}.}
\end{center}
\end{figure*}

Over the wide pressure range from UHV up to 0.1 Torr our {\em in situ} XPS measurements thus reveal no qualitative change in the surface catalytic function. That the pressure gap is nevertheless not fully closed is indicated by the much higher O coverage obtained just at the highest attainable pressure of 1\,Torr. This coverage of $\sim 0.75$\,ML would rather be consistent with the $(5 \times 5)$ oxidic precursor structure characterized in UHV \cite{LundgrenS5}. It could thus well be that the real gap phenomena just occur in the pressure range above 1\,Torr that we can not yet access due to loss of photoelectron intensity. This tantalizing conjecture is indirectly corroborated by the 1p-kMC simulations that we can well run at ambient pressures. The activation temperature resulting from the metal Pd(100)-only model for 1\,atm is as high as 600\,$^\circ$C, with insignificant catalytic activity of the CO-poisoned state at around 300\,$^\circ$C. As this contradicts the known high activity of Pd(100) at these temperatures \cite{logan}, there must be qualitative physics missing in the employed model, which we assign to the formation of oxidic overlayers at the surface. Support for this hypothesis comes from {\em in situ} XPS measurements at 0.5\,Torr and more oxygen-rich feeds. Fig.~\ref{100-400fit} summarizes the results from an experiment with a 4:1 mixture of O$_2$ and CO. The more oxidizing environment decreases the activation temperature to 270$^\circ$C.

More interestingly, the O coverage above the activation temperature now increases to 0.8 ML, as indicative of the formation of the $\sqrt{5}$ surface oxide. This suspicion is confirmed by the O 1$s$ and Pd 3$d$ spectra shown in detail in Fig.~\ref{100-400fit}d). The O 1$s$ spectrum reveals two components at 528.7 and 529.5 eV, and the Pd 3$d$ level has one component at 336.2 eV, shifted 1.3 eV from the bulk component. This leaves little doubt on the presence of the $\sqrt{5}$ oxide \cite{LundgrenS5,Kostelnik}, cf. Fig.~\ref{100-400fit}e), and is again fully consistent with the interpretation of the CO vibrational data by Gao {\em et al.} at these O$_2$:CO ratios \cite{Goodman}.

While at 1:1 CO:O$_2$ ratio we thus cannot reach high enough total pressures to observe formation of oxidic overlayers, this is possible at more O-rich feeds. Extrapolating this view we would thus conclude that the likely termination at technological (near-stoichiometric and ambient) gas-phase conditions is the surface oxide. Validation of this hypothesis through dedicated {\em in situ} techniques is, however, not only a function of increasing their operation range beyond the presently attainable near-ambient regime. Equally important will be to battle the MTLs of which there are clear signatures already visible at the higher end of pressures studied here. While in the present reactor chamber no explicit measurement of the catalytic activity is possible, both the high-temperature metallic Pd(100) and the $\sqrt{5}$ surface oxide, are highly active. In consequence, diffusion limitations in bringing the CO minority species to the active surface lead to a depletion of CO in the gas-phase directly above the catalyst surface \cite{Matera,Matera2}. This is particularly consequential as the residence time of CO at both active phases is extremely short. Using the DFT-derived kinetic parameters we estimate this residence time as the inverse of the sum of all rate constants of desorption and reaction processes and obtain about $9 \cdot 10^{-10}$\,s. In this situation the catalytic function is highly sensitive to the impingement rate of CO molecules to the surface and thus to MTLs that modify the CO pressure profile directly above the active surface. A conclusive answer to the question whether the active phase in CO oxidation over Pd or other late TM catalysts is an O-covered pristine metal or an oxidic overlayer thus dictates new reactor setups that allow to address technological ambient pressures without suffering from MTLs. We note though that the crucial question hereby is maybe not even which phase is the more active one (metal or oxide), but rather which phase is stabilized -- and if the less reactive phase is stabilized, what can be done to stabilize the other and arrive at an improved catalytic function.

In conclusion, our presented {\em in situ} XPS and 1p-kMC data demonstrates that over the entire pressure range from UHV up to 1 Torr the catalytic activity of Pd at near-stoichiometric pressure ratios can be understood in terms of a CO-poisoned inactive state at low temperatures and above the activation temperature by a highly active state composed of Pd(100) with a high coverage of chemisorbed O. Unfortunately, this does not yet close the pressure gap, as formation of oxidic overlayers seems to emerge precisely at the upper edge of the presently accessible total pressure range. For more O-rich feeds, this formation starts at lower total pressures and we can unambiguously detect the formation of the ($\sqrt{5}\times\sqrt{5}$) surface oxide phase known from UHV studies. Both surface oxide and the high-temperature metallic Pd(100) phase are highly active under the probed gas-phase conditions, making the measurements highly sensitive to mass-transfer limitations in the employed reactor. A true closing of the pressure gap will thus not only require efforts in extending atomic-scale electron spectroscopies to higher pressures, but also reactor set-ups that overcome these limitations. However, already the present results demonstrate how the comprehensive insight provided by cutting-edge HP-XPS helps to qualify the relevant pressure range for the gap phenomenon and provide a better founded perspective on the long-standing controversy over the high pressure active phase.

The authors would like to thank the Swedish Research Council, Swedish Foundation for Strategic Research (SSF), the Crafoord foundation, the Knut and Alice Wallenberg foundation and the Anna and Edwin Berger foundation.The work was also supported by the Director, Office of
Science, Office of Basic Energy Sciences, of the US Department of Energy under Contract No. DE-AC02-05CH11231, the German Research Council and the Research Council of Norway (Project No. 138368/V30). The ALS staff is gratefully acknowledged.

%This work was also supported by the Director,Office of Science, Office of Advanced Scientific Computing Research, Office of Basic Energy Sciences,Materials Sciences and Engineering, and Chemical Sciences,Geosciences, and Biosciences Division of the U.S. Department of Energy under Contract No. DE-AC02-05CH11231

\end{document}